\documentstyle[prl,aps,twocolumn,epsf]{revtex}
\begin {document}

\twocolumn[\hsize\textwidth\columnwidth\hsize\csname
@twocolumnfalse\endcsname

\title
{
Sandpile Models of Self-Organized Criticality
}
\author
{
S. S. Manna
}
\address
{
P. M. M. H., \'Ecole Sup\'erieure de Physique et Chimie Industrielles,
10, rue Vauquelin, 75231 Paris Cedex 05 France \\
and\\
Satyendra Nath Bose National Centre for Basic Sciences,
Block-JD, Sector-III, Salt Lake, Calcutta 700 091, India \\
}

\maketitle
\begin{abstract}

Self-Organized Criticality is the emergence of long-ranged
spatio-temporal correlations in non-equilibrium steady states of slowly
driven systems without fine tuning of any control parameter. Sandpiles
were proposed as prototypical examples of self-organized criticality.
However, only some of the laboratory experiments looking for the evidence
of criticality in sandpiles have reported a positive outcome.
On the other hand a large number of theoretical models have been
constructed that do show the existence of such a critical state. We
discuss here some of the theoretical models as well as some experiments.

\end{abstract}

\vskip2pc]

  The concept of Self-Organized Criticality (SOC) was introduced by
Bak, Tang and Wiesenfeld (BTW) in 1987 \cite{btw}. It says that there
is a certain class of systems in nature whose members become critical
under their own dynamical evolutions. An external agency drives the
system by injecting some mass (in other examples, it could be the
slope, energy or even local voids) into it. This starts a transport
process within the system: Whenever the mass at some local region
becomes too large, it is distributed to the neighbourhood by using
some local relaxation rules.  Globally, mass is transported by many
such successive local relaxation events. In the language of sandpiles,
these together constitute a burst of activity called an avalanche. If
we start with an initial uncritical state, initially most of the
avalanches are small, but the range of sizes of avalanches grows with
time. After a long time, the system arrives at a critical state, in
which the avalanches extend over all length and time
scales. Customarily, critical states have measure zero in the phase
space.  However, with self-organizing dynamics, the system finds these
states in polynomial times, irrespective of the initial state
\cite{bakbook,jensen,dharrev}.

  BTW used the example of a sandpile to illustrate their ideas about SOC.  
If a sandpile is formed on a horizontal circular base with any arbitrary
initial distribution of sand grains, a sandpile of fixed conical shape
(steady state) is formed by slowly adding sand grains one after another
(external drive). The surface of the sandpile in the steady state on the
average makes a constant angle known as the angle of repose, with the
horizontal plane. Addition of each sand grain results in some activity on
the surface of the pile: an avalanche of sand mass follows, which
propagates on the surface of the sandpile. Avalanches are of many
different sizes and BTW argued that they would have a power law
distribution in the steady state.

  There are also some other naturally occurring phenomena which are
considered to be examples of SOC. Slow creeping of tectonic plates
against each other results intermittent burst of stress release during
earthquakes. The energy released is known to follow power law
distributions as described by the well known Gutenberg-Richter Law
\cite{guten}. The phenomenon of earthquakes is being studied using SOC
models \cite{earthquake}. River networks have been found to have fractal
properties. Water flow causes erosion in river beds, which in turn
changes the flow distribution in the network. It has been argued that the
evolution of river pattern is a self-organized dynamical process
\cite{river}. Propagation of forest fires \cite{forest} and biological
evolution processes \cite{bio} have also been suggested to be examples of SOC.
\begin{figure}[h]
\centerline{\epsfxsize=245pt\epsfbox{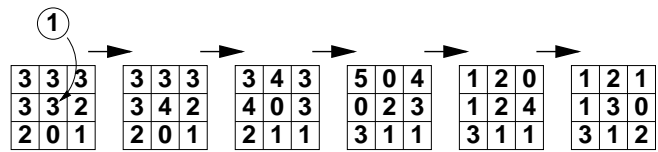}}
\vskip -2.0cm
\caption
{
An avalanche of the Abelian Sandpile Model, generated on a
$3 \times 3$ square lattice.  A sand grain is dropped on a stable
configuration at the central site.  The avalanche created has size $s
=6$, area $a = 6$, life-time $t = 4$ and the radius $r = \sqrt 2$.
}
\end{figure}

  Laboratory experiments on sandpiles, however, have not always found
evidence of criticality in sandpiles. In the first experiment, the
granular material was kept in a semicircular drum which was slowly rotated
about the horizontal axis, thus slowly tilting the free surface of the
pile. Grains fell vertically downward and were allowed to pass through the
plates of a capacitor. Power spectrum analysis of the time series for the
fluctuating capacitance however showed a broad peak, contrary to the
expectation of a power law decay, from the SOC theory \cite{chicago}.

  In a second experiment, sand was slowly dropped on to a horizontal
circular disc, to form a conical pile in the steady state. On further
addition of sand, avalanches were created on the surface of the pile, and
the outflow statistics was recorded. The size of the avalanche was
measured by the amount of sand mass that dropped out of the system. It was
observed that the avalanche size distribution obeys a scaling behaviour
for small piles. For large piles, however, scaling did not work very
well. It was suggested that SOC behavior is seen only for small sizes, and
very large systems would not show SOC \cite{ibm}.

  Another experiment used a pile of rice between two vertical glass plates 
separated by a small gap. Rice grains were slowly dropped on to the pile.
Due to the anisotropy of grains, various packing configurations were observed. 
In the steady state, avalanches of moving rice grains refreshed the surface 
repeatedly. SOC behaviour was observed for grains of large aspect ratio, but 
not for the less elongated grains \cite{oslo}.

  Theoretically, however, a large number of models have been proposed
and studied. Most of these models study the system using cellular
automata where discrete, as well as continuous, variables are used for
the heights of sand columns. Among them, the Abelian Sandpile model
is most popular \cite{btw,asm}. Other models of self organized
criticality have been studied but will not be discussed here. These
include the Zheng model which has modified rules for sandpile
evolution \cite{zheng}, a model for Abelian distributed processors and other
stochastic rule models \cite{dharrev}, the Eulerian
Walkers model \cite{pdkd} and the Takayasu aggregation model
\cite{takayasu}. 

In the Abelian sandpile model, we associate a non-negative integer
variable $h$ representing the height of the `sand column' 
with every lattice site on a $d$-dimensional lattice (in general on
any connected graph). One often starts with an arbitrary
initial distribution of heights. Grains are added one at a time at
randomly selected sites ${\cal O}$: $h_{\cal O}\rightarrow h_{\cal
O}+1$. The sand column at any arbitrary site $i$ becomes unstable when
$h_i$ exceeds a previously selected threshold value $h_c$ for the
stability. Without loss of generality, one usually chooses
$h_c=2d-1$. An unstable sand column always topples.  In a toppling,
the height is reduced as: $h_i \rightarrow h_i-2d$ and all the $2d$
neighbouring sites $\{j\}$ gain a unit sand grain each: $h_j
\rightarrow h_j+1$.  This toppling may make some of the neighbouring
sites unstable. Consequently, these sites will topple again, possibly
making further neighbours unstable. In this way a cascade of topplings
propagates, which finally terminates when all sites in the system
become stable (Fig. 1). One waits until this avalanche stops before adding the
next grain. This is equivalent to assuming that the rate of adding
sand is much slower than the natural rate of relaxation of the
system. The wide separation of the `time scale of drive' and `time
scale of relaxation' is common in many models of SOC. For instance, in
earthquakes, the drive is the slow tectonic movement of continental
plates, which occurs over a timescale of centuries, while the actual
stress relaxation occurs in quakes, whose duration is only a few
seconds.  This separation of time scales is usually considered to be a
defining characteristic of SOC. However, Dhar has argued that the wide
separation of time scales should not be considered as a necessary
condition for SOC in general \cite{dharrev}. Finally, the system must
have an outlet, through which the grains go out of the system, which
is absolutely necessary to attain a steady state. Most popularly, the
outlet is chosen as the $(d-1)$ dimensional surface of a
$d$-dimensional hypercubic system.
\begin{figure}[h]
\vskip 0.5 cm
\centerline{\epsfxsize=245pt\epsfbox{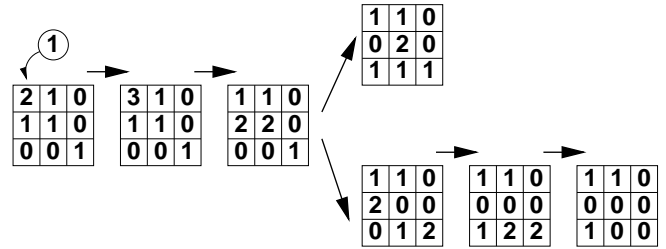}}
\caption
{
An example to show that a directed slope model is
non-Abelian.  Two slopes are measured from any site $(i,j)$ as
$h(i,j)-h(i,j+1)$ and $h(i,j) - h(i+1,j+1)$.  If either of them is
greater than 1, two grains are transferred from $(i,j)$ and are given
one each to $(i,j+1)$ and $(i+1,j+1)$.  On dropping a grain on the
initial stable configuration, we see that finally two different height
configurations result due to two different sequences of topplings
[20].
}
\end{figure}

  The beauty of the Abelian model is that the final stable height
configuration of the system is independent of the sequence in which
sand grains are added to the system to reach this stable configuration
\cite{asm}. On a stable configuration $\cal C$, if two grains are
added, first at $i$ and then at $j$, the resulting stable
configuration $\cal C'$ is exactly same in case the grains were added
first at $j$ and then at $i$. In other sandpile models, where the
stability of a sand column depends on the local slope or the local
Laplacian, the dynamics is not Abelian, since toppling of one unstable site
may convert another unstable site to a stable site (Fig. 2). Many such
rules have been studied in the literature \cite{mannaphysica,kada}.

  An avalanche is a cascade of topplings of a number of sites created
on the addition of a sand grain. The strength of an avalanche in
general, is a measure of the effect of the external perturbation
created due to the addition of the sand grain.  Quantitatively, the
strength of an avalanche is estimated in four different ways: (i) size
$(s)$: the total number topplings in the avalanche, (ii) area $(a)$:
the number of distinct sites which toppled, (iii) life-time $(t)$: the
duration of the avalanche and (iv) radius $(r)$: the maximum distance
of a toppled site from the origin. These four different
quantities are not independent and are related to each other by scaling laws.
Between any two measures $x,y \in \{s,a,t,r\}$ one can
define a mutual dependence as: $ <y> \sim x^{\gamma_{xy}}.  $ These
exponents are related to one another, e.g., $\gamma_{ts} = \gamma_{tr}
\gamma_{rs}$.  For the ASM, it can be proved that the avalanche clusters
cannot have any holes. It has been shown that $\gamma_{rs} = 2$ in
two dimensions. It has also benn proved that $\gamma_{rt}$ = 5/4 \cite{burn}.
A better way to estimate the $\gamma_{tx}$ exponents is to average
over the intermediate values of the size, area and radius at every
intermediate time step during the growth of the avalanche.

  Quite generally, the finite size scaling form for the probability
distribution function for any measure $x \in \{s,a,t,r\}$ is taken to be:
\[
P(x) \sim x^{-\tau_x} f_x \left (\frac {x}{L^{\sigma_x}}\right ). \quad
\]
The exponent $\sigma_x$ determines the variation of the cut-off of the
quantity $x$ with the system size $L$. Alternatively, sometimes it is
helpful to consider the cumulative probability distribution $F(x) =
$$\int^{L^{\sigma_x}}_x P(x)dx$ which varies as
$x^{1-\tau_x}$. However, in the case of $\tau_x = 1$, the variation
should be in the form $F(x) = C -$ log$(x)$. Between any two measures,
scaling relations like $\gamma_{xy}=(\tau_x-1)/(\tau_y-1)$
exist. Recently, the scaling assumptions for the avalanche sizes have
been questioned. It has been argued that there actually exists a
multifractal distribution instead \cite{stella}.

  Numerical estimation for the exponents have yielded scattered values. For
example estimates of the exponent $\tau_s$ range from 1.20 \cite
{mannaphysica} to 1.27 \cite{chessa} and 1.29 \cite{lubeck}.

  We will now look into the structure of avalanches in more detail.  A
site $i$ can topple more than once in the same avalanche. The set of
its neighbouring sites $\{j\}$, can be divided into two subsets.
Except at the origin $\cal O$, where a grain is added from the outside, for a toppling,
the site $i$ must receive some grains from some of the neighbouring
sites $\{j_1\}$ to exceed the threshold $h_c$. These sites must have
toppled before the site $i$. When the site $i$ topples, it loses $2d$
grains to the neighbours, by giving back the grains it has received
from $\{j_1\}$, and also donating grains to the other  neighbours
$\{j_2\}$. Some of these neighbours may topple later, which returns
grains to the site $i$ and its height $h_i$ is raised. The following
possibilities may arise: (i) some sites of $\{j_2\}$ may not topple at
all; then the site $i$ will never re-topple and is a singly toppled
site on the surface of the avalanche.  (ii) all sites in $\{j_2\}$
topple, but no site in $\{j_1\}$ topples again; then $i$ will be a
singly toppled site, surrounded by singly toppled sites.  (iii) all
sites in $\{j_2\}$ topple, and some sites of $\{j_1\}$ re-topple;
then $i$ will remain a singly toppled site, adjacent to the doubly
toppled sites.  (iv) all sites in $\{j_2\}$ topple, and all sites of
$\{j_1\}$ re-topples; then the site $i$ must be a doubly toppled site.
This implies that the set of at least doubly toppled sites must be
surrounded by the set of singly toppled sites. Arguing in a similar
way will reveal that sites which toppled at least $n$ times, must be a
subset and also are surrounded by the set of sites which toppled at
least $(n-1)$ times. Finally, there will be a central region in the
avalanche, where all sites have toppled a maximum of $m$ times. The origin
of the avalanche $\cal O$, where the sand grain was dropped, must be a
site in this maximum toppled zone. Also the origin must be at the
boundary of this $m^{\rm th}$ zone, since otherwise it should have toppled
$(m+1)$ times \cite{G-M}.
\begin{figure}[h]
\vskip -3.8 cm
\centerline{\epsfxsize=245pt\epsfbox{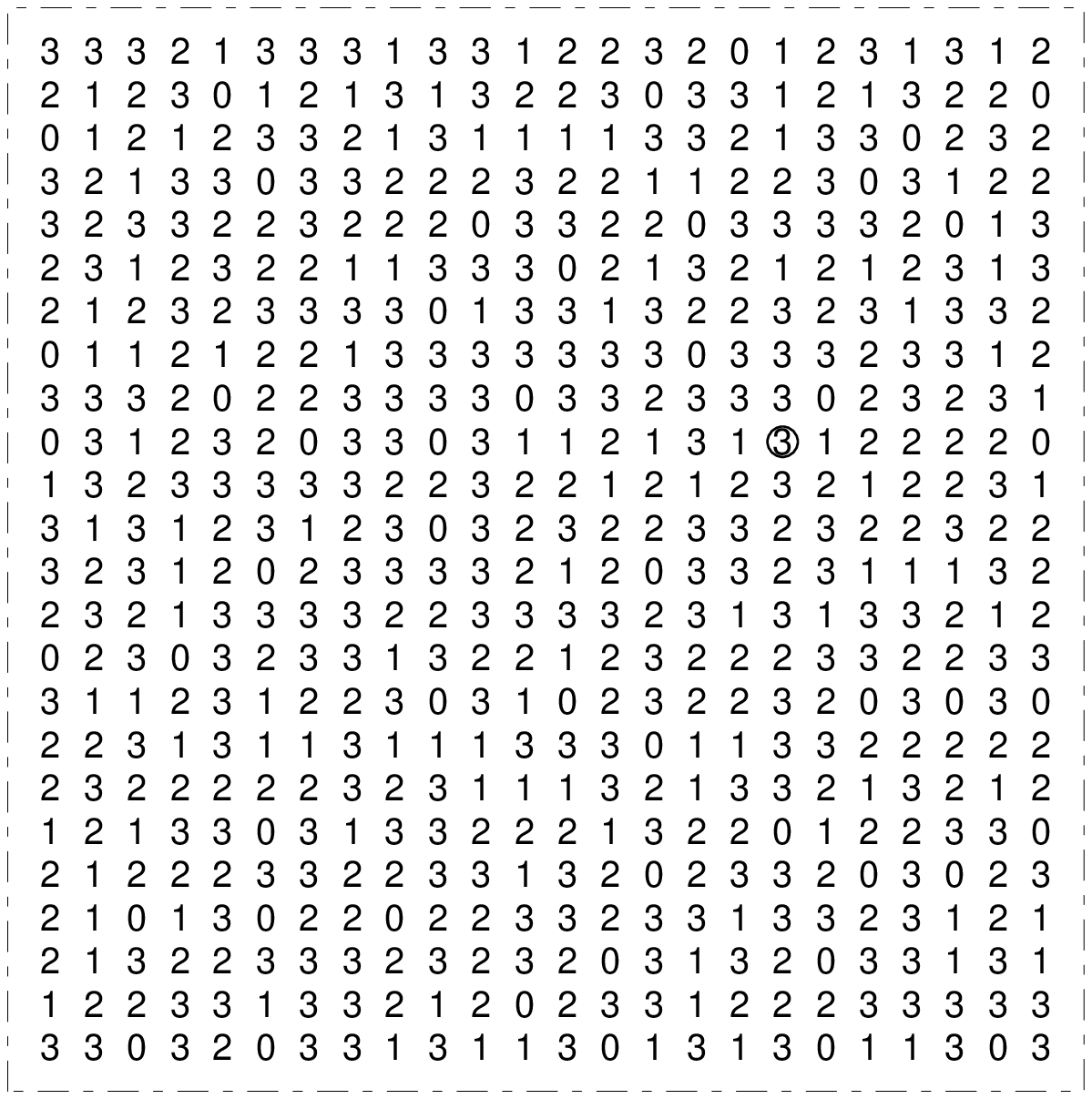}}
\vskip -5.3 cm
\centerline{\epsfxsize=245pt\epsfbox{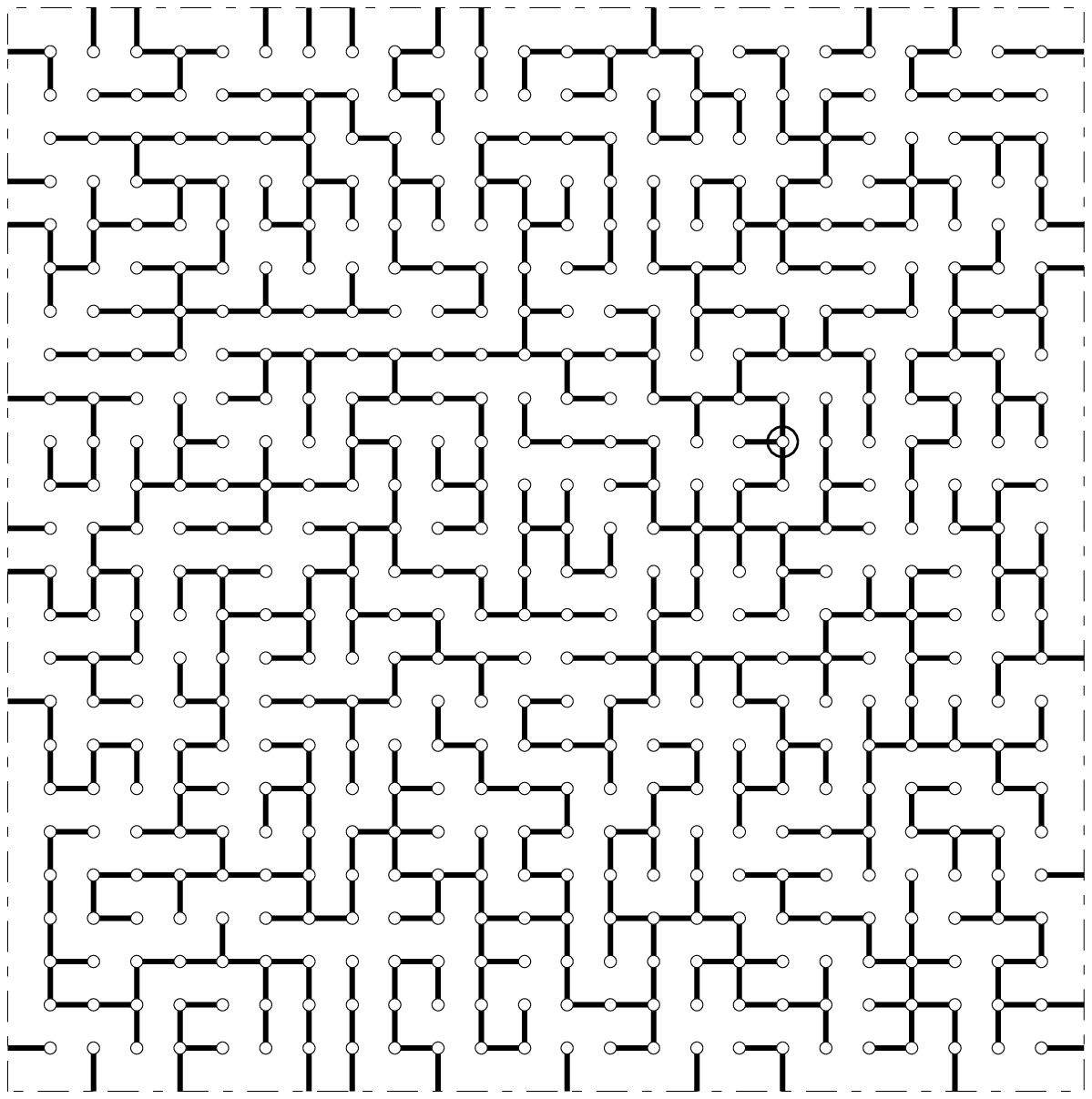}}
\vskip -5.3 cm
\centerline{\epsfxsize=245pt\epsfbox{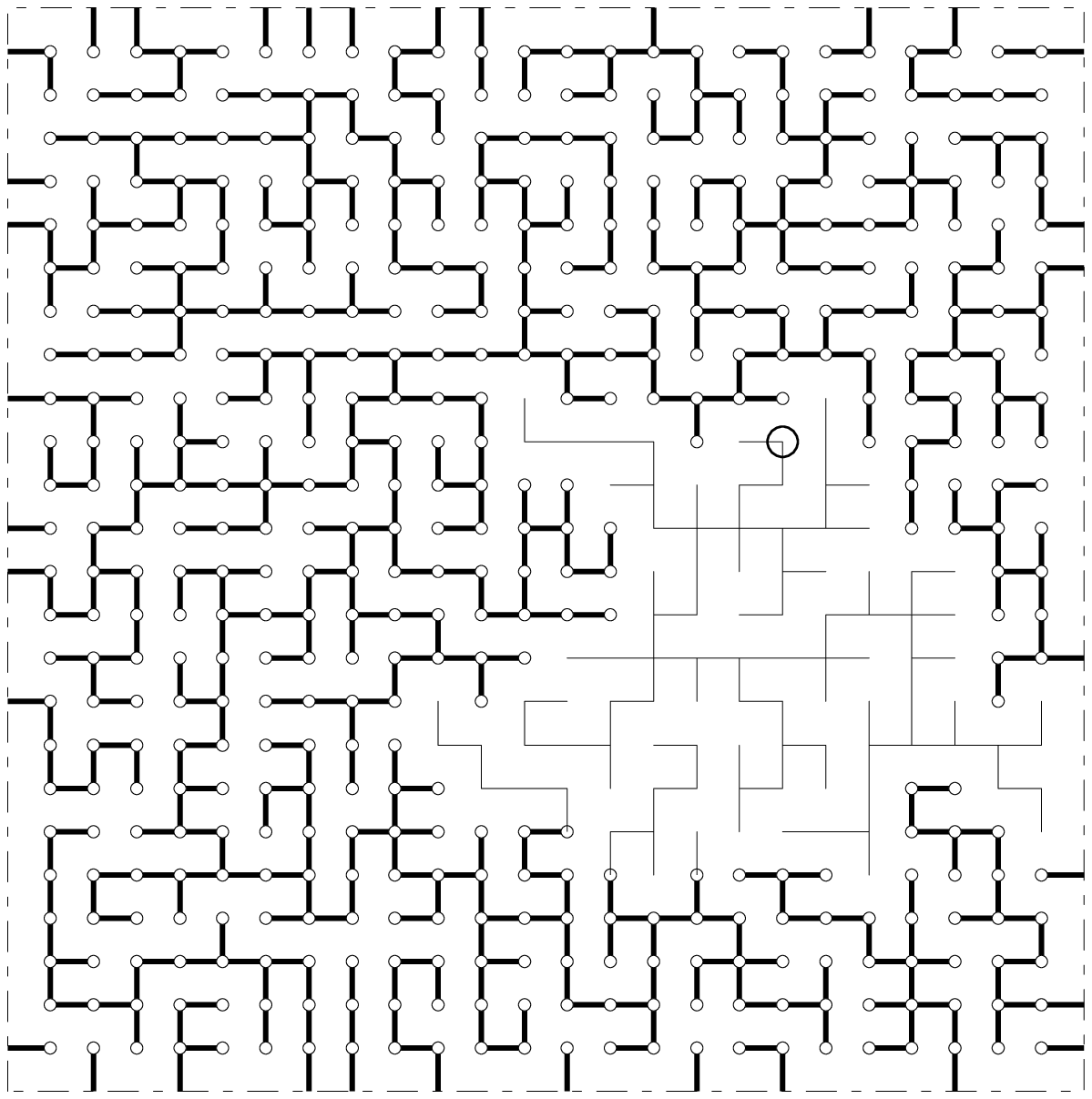}}
\vskip -2.0cm
\caption
{
(a) An example of the height distribution in a recurrent
configuration ${\cal C'}$ on a $24 \times 24$ square lattice. This
configuration is obtained by dropping a grain a some previous
configuration ${\cal C}$ at the encircled site. (b) The spanning tree
representation of the configuration ${\cal C'}$. (c) A new
configuration ${\cal C''}$ is obtained by taking out one grain at the
encircled site from the configuration ${\cal C'}$. A spanning tree
cannot be obtained for ${\cal C''}$.  The bonds of the spanning tree
corresonding to the forbidden sub-configuration in ${\cal C''}$ are shown by the thin lines.
}
\end{figure}

  Using this idea, we see that the boundary sites on any arbitrary
system can topple at most once in any arbitrary number of
avalanches. Similar restrictions are true for inner sites also. A
$(2n+1) \times (2n+1)$ square lattice can be divided into $(n+1)$
subsets which are concentric squares. Sites on the $m$-th such square
from the boundary can topple at most $m$ times, where as the central
site cannot topple more than $n$ times in any avalanche.

  Avalanches can also be decomposed in a different way, using 
{\it Waves of Toppling}. Suppose, on a stable configuration $\cal C$,
a sand grain is added at the site ${\cal O}$. The site is toppled
once, but is not allowed to topple for the second time, till all other
sites become stable. This is called the first wave. It may happen that
after the first wave, the site ${\cal O}$ is stable; in that case the
avalanche has terminated. If the site ${\cal O}$ is still unstable it
is toppled for the second time, and all other sites are allowed to
become stable again; this is called the second wave, and so on. It
was shown, that in a sample where all waves occur with equal weights,
the probability of occurrence of a wave of area $a$ is $D(a) \sim 1/a$
\cite{priezzhev}.

  It is known that the stable height configurations in ASM are of two
types: {\it Recurrent} configurations appear only in the steady state
with uniform probabilities, whereas  {\it Transient}
configurations occur in the steady state with zero
probability. Since long range correlations appear only in the steady
states, it implies that the recurrent configurations are
correlated. This correlation is manifested by the fact that certain
clusters of connected sites with some specific distributions of
heights never appear in any recurrent configuration. Such clusters are
called the forbidden sub-configurations.  It is easy to show
that two zero heights at the neighbouring sites: {\bf \sf (0$-$0)} or,
an unit height with two zero heights at its two sides: {\bf \sf
(0$-$1$-$0)} never occur in the steady state. There are also many more
forbidden sub-configurations of bigger sizes.

  An $L \times L$ lattice is a graph, which has all the sites and all
the nearest neighbour edges (bonds). A {\it Spanning Tree} is a
sub-graph of such a graph, having all sites and some bonds. It has no
loop and therefore, between any pair of sites there exists an unique
path through a sequence of bonds. There can be many possible Spanning
trees on a lattice.  These trees have interesting statistics in a
sample where they are equally likely.  Suppose when we randomly select
such a tree and then randomly select one of the unoccupied bonds and
occupy it, it forms a loop of length $\ell$. It has been shown that
these loops have the length distribution $D(\ell) \sim \ell^{-8/5}$.
Similarly, if a bond of a Spanning tree is randomly selected and
deleted, then it divides into two fragments. The sizes of the two
fragments generated follow a probability distribution $D(a) \sim
a^{-11/8}$ \cite{span}. It was also shown that every recurrent
configuration of the Abelian model on an arbitrary lattice has a
one-to-one correspondence to a random Spanning tree graph on the same
lattice. Therefore, there are exactly the same number of distinct
Spanning trees as the number of recurrent Abelian sandpile model
configurations on any arbitrary lattice \cite{burn}.  Given a stable
height configuration, there exists an unique prescription to obtain
the equivalent Spanning tree. This is called the {\it Burning} method
\cite{burn}. A fire front, initially at every site outside the
boundary, gradually penetrates (burns) into the system using a
deterministic rule. The paths of the fire front constitute the
Spanning tree. A fully burnt system is recurrent, otherwise it is transient.

  Suppose, addition of a grain at the site ${\cal O}$ of a stable
recurrent configuration $\cal C$, leads to another stable
configuration $\cal C'$. Is it possible to get back the configuration
$\cal C$ knowing $\cal C'$ and the position of $\cal O$? This is done
by {\it Inverse toppling} \cite{inverse}. Since $\cal C'$ is
recurrent, a corresponding Spanning tree ($\cal C'$) exists.  Now, one grain at
$\cal O$ is taken out from $\cal C'$ and the configuration ${\cal
C''}$= ${\cal C'}-\delta_{{\cal O}j}$ is obtained. This means on
ST($\cal C'$), one bond is deleted at ${\cal O}$ and it is divided
into two fragments. Therefore one cannot burn the configuration $\cal
C''$ completely since the resulting tree has a hole consisting of at
least the sites of the smaller fragment (Fig.3). This implies that $\cal C''$
has a forbidden sub-configuration $(F_1)$ of equal size and $\cal C''$ is not recurrent. On
$(F_1)$, one runs the inverse toppling process: 4 grains are added to
each site $i$, and one grain each is taken out from all its neighbours
$\{j\}$. The cluster of $f_1$ sites in $F_1$ is called the first
inverse avalanche. The lattice is burnt again. If it still has a
forbidden sub-configuration 
($F_2$), another inverse toppling process is executed, and is called
the second inverse avalanche. The size of the avalanche is: $s = f_1 +
f_2 + f_3 + .... $, and the $f_1$ is related to the maximum toppled
zone of the avalanche. From the statistics of random spanning trees
\cite{span} it is clear that $f_1$ should have the same statistics of
the two fragments of the tree generated on deleting one bond. Therefore
the maximum toppled zone also has a power law distribution of the size,
$D(a) \sim a^{-11/8}$.

  Sandpile models with stochastic evolution rules have also been
studied. The simplest of these is a {\it Two-state} sandpile model, A
stable configuration of this system consists of sites, either vacant
or occupied by at most one grain. If there are two or more grains at a
site at the same time we say there is a {\it collision}. In this case,
all grains at that site are moved. Each grain chooses a randomly
selected site from the neighbours and is moved to that site. The
avalanche size is the total number of collisions in an avalanche. From
the numerical simulations, the distribution of avalanche sizes is found to
follow a power law, characterized by an exponent $\tau_s \approx 1.27$
\cite{two-state}. This two-state model has a nontrivial dynamics even
in one dimension \cite{sneppen}. Recently, it has been shown that
instead of moving all grains, if only two grains are moved randomly
leaving others at the site, the dynamics is Abelian \cite{dhar}.

  Some other stochastic models also have nontrivial critical behaviour
in one dimension. To model the dynamics of rice piles, Christensen
et. al.  studied the following slope model \cite{ricemodel}. On a
one-dimensional lattice of length $L$, non-negative integer variable
$h_i$ represents the height of the sand column at the site $i$. The
local slope $z_i=h_i - h_{i+1}$ is defined, maintaining zero height on
the right boundary.  Grains are added only at the left boundary
$i=1$. Addition of one grain $h_i \rightarrow h_i +1 $ implies an
increase in the slope $z_i \rightarrow z_i +1 $. If at any site, the
local slope exceeeds a pre-assigned threshold value $z_i^c$, one grain
is transferred from the column at $i$ to the column at $(i+1)$. This
implies a change in the local slope as: $z_i \rightarrow z_i -2$ and
$z_{i\pm 1} \rightarrow z_{i\pm 1} +1 $.  The thresholds of the
instability $z_i^c$ are dynamical variables and are randomly chosen
between 1 and 2 in each toppling. Numerically, the avalanche sizes are
found to follow a power law distribution with an exponent $\tau_s
\approx 1.55$ and the cutoff exponent was found to be $\sigma_s
\approx 2.25$. This model is referred as the Oslo model.

  Addition of one grain at a time, and allowing the system to relax to
its stable state, implies a zero rate of driving of the system. What
happens when the driving rate is finite? Corral and Paczuski studied
the Oslo model in the situation of nonzero flow rate. Grains were
added with a rate $r$, i.e., at every (1/$r$) time updates, one grain
is dropped at the left boundary $i=1$. They observed a dynamical
transition separating intermittent and continuous flows \cite{corral}.

Many different versions of the sandpile model have been studied.
However the precise classification of various models in different
universality classes in terms of their critical exponents is not yet
available and still attracts much attention \cite{mannaphysica,kada}.
Exact values of the critical exponents of the most widely studied
Abelian model 
are still not known in two dimensions. Some effort has also been made 
towards the analytical calculation of avalanche size exponents
\cite{ktitarev,boettcher,kti-pri}.  Numerical studies for these
exponents are found to give scattered values.  On the other hand the
two-state sandpile model is believed to be better behaved and there is
good agreement of numerical values of its exponents by different
investigators.  However, whether the Abelian model and the two-state model
belong to the same universality class or not is still an unsettled
question \cite{vesp}.

  If a real sandpile is to be modeled in terms of any of these
sandpile models or their modifications, it must be a slope model,
rather than a height model.  However, not much work has been done to
study the slope models of sandpiles \cite{mannaphysica,kada}.  Another
old question is whether the conservation of the grain number in the
toppling rules is a necessary condition to obtain a critical
state.  It has been shown already that too much non-conservation leads
to avalanches of characteristic sizes \cite{mkk}. However, if grains
are taken out of the system slowly, the system is found to be critical
in some situations. A non-conservative version of the Abelian sandpile
model with
directional bias shows a mean field type critical behaviour
\cite{mcc}. Therefore, the detailed role of the conservation of the
grain numbers during the topplings is still an open question.

  We acknowledge D. Dhar with thanks for a critical reading of the
manuscript and for useful comments.

Electronic address: manna@boson.bose.res.in

\end {document}